\documentclass[twocolumn,showpacs,prb,floatfix]{revtex4}
\usepackage{epsfig}
\usepackage{tabularx}

\newcommand{\be}{\begin{equation}}
\newcommand{\ee}{\end{equation}}
\newcommand{\beqn}{\begin{eqnarray}}
\newcommand{\eeqn}{\end{eqnarray}}
\newcommand{\nsw}{N_{\mathrm{sweep}}}
\newcommand{\nsa}{N_{\mathrm{samp}}}

\newcommand{\chisg}{\chi_{_{\mathrm{SG}}}}

\begin{document}

\title{Numerical studies of the two- and three-dimensional gauge 
glass\\at low temperature}
\author{Helmut G.~Katzgraber}
\affiliation{Department of Physics, University of
California, Davis, California 95616}

\author{A.~P.~Young}
\email{peter@bartok.ucsc.edu}
\homepage{http://bartok.ucsc.edu/peter}
\altaffiliation{Temporary address:
Department of Theoretical Physics, 1, Keble Road, Oxford OX1 3NP,
England}
\affiliation{Department of Physics,
University of California,
Santa Cruz, California 95064}

\date{\today}

\begin{abstract}
We present results from Monte Carlo simulations of the two- and
three-dimensional gauge glass at low temperature
using the parallel tempering Monte Carlo method. Our results in two dimensions
strongly support the transition being at $T_c=0$. A finite-size scaling
analysis, which works well only for the larger sizes and lower temperatures,
gives the stiffness exponent $\theta = -0.39 \pm 0.03$. In three dimensions
we find $\theta = 0.27 \pm 0.01$, compatible with recent results from domain
wall renormalization group studies.
\end{abstract}

\pacs{75.50.Lk, 75.40.Mg, 05.50.+q}
\maketitle

\section{Introduction}
\label{introduction}

The gauge glass is a model which is often used to describe the vortex glass
transition in high-temperature superconductors. Consequently there has been a
substantial amount of theoretical work attempting to understand its
properties.  Nonetheless, there are still areas of disagreement.  For example,
in two dimensions there is an ongoing controversy as to whether a spin-glass
transition occurs at finite temperature or not. In addition, in three
dimensions, although it is known that the system exhibits a finite-temperature
transition, see, e.g., Ref.~\onlinecite{olson:00}, there is no consensus on the
value of the stiffness exponent.

Recent results by means of resistively-shunted junction (RSJ) dynamics
by Kim\cite{kim:00} claim evidence for
a finite-temperature transition
in two dimensions with a transition temperature $T_c = 0.22$. 
Choi and Park\cite{choi:99} find the same value for $T_c$
by studying the scaling of the spin-glass susceptibility via Monte Carlo
simulations.
These claims for a finite $T_c$ are in contrast to
results by Granato\cite{granato:98} and Hyman {\em et al.}\cite{hyman:95}
who also use RSJ dynamics and find evidence of a zero-$T$ transition.
In addition, Monte Carlo simulations by Fisher {\em et al.}\cite{fisher:91}
and Reger and Young\cite{reger:93} show evidence of a zero-temperature
transition, although the simulations were not taken down to extremely low
temperatures. 

In this work we perform Monte Carlo simulations of the two-dimensional
gauge glass, using parallel tempering~\cite{hukushima:96,marinari:98b} to go
to significantly lower temperatures than was possible in earlier work. In
particular we are now able to cover the temperature range where the claimed
spin-glass transition\cite{choi:99,kim:00} takes place. We find strong
evidence that $T_c=0$.

In addition, we study the gauge glass in three dimensions at very
low, but finite, temperatures to
provide a good estimate of the stiffness exponent $\theta$. The motivation for
this is that earlier estimates, which were obtained from
ground state methods, are inconsistent.
One group\cite{reger:91,gingras:92,kosterlitz:97,maucourt:97} finds values
consistent with $\theta \approx 0$, whereas 
another\cite{akino:02,cieplak:92,moore:94,kosterlitz:98} finds
$\theta$ in the range $0.26$ -- $0.31$.
Ground state methods involve computing the ground state with two
different boundary conditions, and one of the main reasons for the large
range of estimates for $\theta$ is that different groups made
different assumptions about the
optimal form of boundary
condition changes, see, e.g., Ref.~\onlinecite{akino:02}. This difficulty
is avoided in the
finite-$T$ Monte Carlo approach adopted here. We find that
$\theta = 0.27 \pm 0.01$, which agrees with the results of the
second group of ground state calculations mentioned above. 

In Sec.~\ref{model-observables} we introduce the model and observables as well
as details regarding the Monte Carlo technique used. Some results from
finite-size scaling are discussed in Sec.~\ref{finite-size-scaling}. 
Our results in
two and three dimensions are presented in Secs.~\ref{results:2d} and
\ref{results:3d}, respectively, and the conclusions are summarized in
Sec.~\ref{conclusions}.

\section{Model, Observables And Equilibration}
\label{model-observables}

The Hamiltonian of the gauge glass is given by
\begin{equation}
{\cal H} = -J \sum_{\langle i, j\rangle} \cos(\phi_i - \phi_j - A_{ij}),
\label{hamiltonian}
\end{equation}
where the sum ranges over nearest neighbors on a square lattice in $D$ 
dimensions of size $N = L^D$ and $\phi_i$ represent
the angles of the $XY$ spins. Periodic boundary conditions are applied.
The $A_{ij}$ are quenched random variables uniformly distributed between 
$[0,2\pi]$.\cite{katzgraber:01a} Because $A_{ij}$ represent the line
integral of the vector potential between sites $i$ and $j$, we have the
constraint that $A_{ij} = - A_{ji}$. In this work we set $J = 1$.

Traditionally one uses the Binder ratio\cite{binder:81} to estimate
the critical temperature $T_c$. In this method one plots the ratio of the fourth 
moment and the second moment squared of the spin-glass order parameter as
a function of temperature for different system sizes $L$. The crossing
point of all curves identifies $T_c$. For the gauge glass
the Binder ratio cannot exceed unity,\cite{olson:00} thus the splaying of the
curves is small and $T_c$ is difficult to establish.

In order to avoid this
problem we use a method introduced by Reger and Young\cite{reger:91} in which
one calculates the current $I$ defined as the derivative of the free energy $F$
with respect to an infinitesimal twist to the boundaries, i.e.,
\begin{equation}
I(L) = \frac{1}{L} \sum_i \sin(\phi_i - \phi_{i+\hat{x}} - 
A_{i\,i+\hat{x}}) \; .
\end{equation}
In this case the twist is applied along the $\hat{x}$ direction. 
As the gauge fields $A_{ij}$ are uniformly distributed we 
expect $[\langle I(L)\rangle ]_{\rm av} = 0$,
where $[\cdots]_{\rm av}$ represents an average over disorder
and $\langle \cdots \rangle$ represents a thermal average. 
We calculate the root-mean-square current
\begin{equation}
I_{\rm rms} = \sqrt{[\langle I_\alpha(L)\rangle \langle I_\beta(L)\rangle]_{\rm av}} \; .
\end{equation}
Here $\alpha$, $\beta$ denote two replicas of the system with the {\em same} 
disorder introduced to avoid any bias in the calculation of the average
of the square of the currents.
Unlike the Binder ratio, the currents are not restricted to converge 
to unity for $T \rightarrow 0$, thus any potential splaying of the data is 
easily visible. In fact, for $T < T_c$ the currents scale as $I_{\rm rms} \sim
L^\theta$ where $\theta$ is the (positive) stiffness
exponent, see Refs.~\onlinecite{fisher:87,huse:87,fisher:88} for a
general discussion of the stiffness exponent in the context of the ``droplet
model'' of spin glasses, and also
Ref.~\onlinecite{fisher:91b}
which discusses the stiffness exponent specifically in the context
of the vortex glass.

Note that the stiffness exponent is also often defined in terms of the free
energy change for a {\em finite}\/ twist, by $\pi$ for example. We are
assuming here that any
reasonable measure of the stiffness will give the same exponent $\theta$,
whether it be the free energy change due to a finite twist or, as here, the
derivative with respect to an infinitesimal twist. This is consistent with the
usual assumptions of universality for phenomena on long length scales, that
the same exponent will be obtained for the relation between, say, energy and
length scale, for any reasonable definition of energy.
Presumably a plot of the
free energy against twist angle $\Theta$ will have quite complicated
structure,
with local regions varying quadratically as a function $\Theta$ and
proportional to $L^{d-2}$ (where only spin-wave excitations are generated),
but with ``kinks''
where which the system changes
from one parabola to another
due to a vortex changing position. Hence, in effect, we are
assuming that the typical magnitude of
the slope of this curve is of the same order in $L$ as the range of its
variation with $\Theta$. This requires that the typical
spacing between kinks
$\Delta \Theta$ satisfies $L^\theta \sim (\Delta\Theta)^2 L^{d-2}$,
i.e.  $\Delta\Theta \sim L^{-(d - 2 - \theta)/2}$.

For a finite-temperature transition, $T_c > 0$, we expect $\theta > 0$,
since then the ordered state at $T=0$ is ``stiff'' on large scales,
and so will presumably
resist small thermal fluctuations.  On the other hand
for a zero-temperature transition, we expect $\theta < 0$, because the
system will then easily break up under the influence of thermal fluctuations. 

In order to compare with the results of Choi and Park\cite{choi:99} we have
also calculated the spin-glass susceptibility $\chisg$ defined by
\begin{equation}
\chisg = N[\langle q^2\rangle]_{\rm av} \; ,
\label{chisg}
\end{equation}
where $q$ is the spin-glass order parameter defined by
\begin{equation}
q = \frac{1}{N} \sum_j^N \exp[i(\phi_j^\alpha - \phi_j^\beta)]
\end{equation}
and $\alpha$, $\beta$ are two replicas of the system with the same
disorder.
This susceptibility diverges at the transition. However, as we shall see in
Sec.~\ref{finite-size-scaling}, its finite-size scaling behavior is more
complicated than that of $I_{\rm rms}$.

\begin{figure}
\centerline{\epsfxsize=\columnwidth \epsfbox{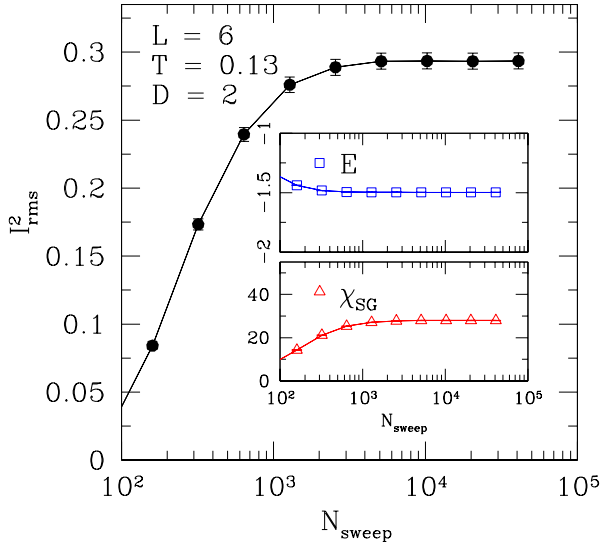}}
\vspace{-1.0cm}
\caption{
Average current squared $I_{\rm rms}^2$, energy $E$
and spin-glass susceptibility $\chisg$
as a function of Monte Carlo sweeps $\nsw$, that each of the replicas perform, 
averaged over the last half of the sweeps. Note that the different observables
equilibrate roughly at the same time and appear to be independent
of the number of sweeps. The data shown is for $D = 2$, $L = 6$ and $T =
0.13$, the lowest temperature studied in two dimensions.
}
\label{equil}
\end{figure}

For the simulations we use the parallel tempering Monte Carlo
method\cite{hukushima:96,marinari:98b} as it allows us to study 
systems at lower temperatures than with conventional methods. 
Because the equilibration test
for short-range spin glasses introduced by Katzgraber {\em et al.}\cite{katzgraber:01}
does not work for the gauge glass since the disorder is not Gaussian,
we test equilibration by the traditional technique of
requiring that different observables are 
independent of the number of Monte Carlo steps $\nsw$. Figure \ref{equil}
shows data in 2D for the energy $E$, spin-glass
susceptibility $\chisg$, as well as the
average current squared $I_{\rm rms}^2$ as a function of Monte Carlo steps. 
One can clearly
see that the different observables saturate at the {\em same} equilibration
time. We show data for an intermediate size
as it allows us to better
illustrate the procedure by simulating with
much longer runs than are necessary to equilibrate. For large sizes
we equilibrate doubling the number of Monte Carlo sweeps between each
measurement until the last three agree within error bars.

We also require that the acceptance ratios for the global moves which
interchange the different temperatures in the parallel tempering scheme
be greater than $0.3$ on average and roughly constant as a function of
temperature.
The equilibration tests for the 3D data have been described
elsewhere.\cite{katzgraber:01a}

Tables \ref{simparams} and \ref{simparams3} show the number of samples $\nsa$ and the 
number of Monte Carlo sweeps $\nsw$ performed by each replica for
each lattice size for $D = 2$ and $D = 3$, respectively. 
\begin{table}
\caption{
\label{simparams}
Parameters of the simulation in two dimensions. $\nsa$ is the
number of samples, i.e., sets of disorder realizations, $\nsw$ is the total
number 
of sweeps simulated for each of the $2 N_T$ replicas for a single sample,
and $N_T$ is the number of temperatures used in the parallel tempering method.
For $L \le 16$ the lowest temperature is $0.13$, while for $L = 24$ 
it is $0.20$.
}
\begin{tabular*}{\columnwidth}{@{\extracolsep{\fill}} c r r l }
\hline
\hline
$L$  &  $\nsa$  & $\nsw$ & $N_T$  \\ 
\hline
4  & 10400 & $8.0 \times 10^4$ & 30 \\
6  & 10150 & $8.0 \times 10^4$ & 30 \\
8  &  8495 & $2.0 \times 10^5$ & 30 \\
12 &  6890 & $8.0 \times 10^5$ & 30 \\
16 &  2500 & $2.0 \times 10^6$ & 30 \\
24 &  2166 & $2.0 \times 10^6$ & 24 \\
\hline
\hline
\end{tabular*}
\end{table}
\begin{table}
\caption{
\label{simparams3}
Parameters of the simulation in three dimensions. The lowest
temperature studied is $T = 0.05$.}
\begin{tabular*}{\columnwidth}{@{\extracolsep{\fill}} c r r l }
\hline
\hline
$L$  &  $\nsa$  & $\nsw$ & $N_T$  \\ 
\hline
3 & 10000 & $6.0 \times 10^3$ &   53 \\
4 & 10000 & $2.0 \times 10^4$ &   53 \\
5 & 10000 & $6.0 \times 10^4$ &   53 \\
6 & 5000  & $2.0 \times 10^5$ &   53 \\
8 & 2000  & $1.2 \times 10^6$ &   53 \\
\hline
\hline
\end{tabular*}
\end{table}
In two dimensions the highest temperature is $1.058$, whereas the
lowest temperature is $0.13$ (for $L = 24$ the lowest temperature is $0.20$).
The number of temperatures $N_T$ is chosen to give satisfactory 
acceptance ratios for the Monte Carlo moves between the temperatures. 
In three dimensions the lowest 
temperature studied is $0.05$ [to be compared with $T_c \approx
0.45$, (Ref.~\onlinecite{olson:00})] whereas the highest temperature is $0.947$.

Because the gauge glass has a vector order parameter symmetry, to speed up the
simulations we discretize the angles of the spins\cite{katzgraber:01a,cieplak:92} 
to $N_{\phi} = 512$. 
To ensure a reasonable acceptance ratio for single-spin
Monte Carlo moves, we pick the proposed new angle for a spin
within a temperature-dependent acceptance window about the current angle.
By tuning a numerical prefactor we ensure the acceptance ratios 
for these local moves are not smaller than 0.2 for each system size at the
lowest temperature simulated.

\section{Finite-Size Scaling}
\label{finite-size-scaling}
We review briefly some salient aspects of finite-size scaling theory, as
applied to the present situation, paying particular attention to features
which are special to a $T=0$ transition. Consider first
the singular part of the {\em bulk}\/ free
energy per spin divided by $T$, a quantity with dimension
$($length$)$$^{-D}$, where $D$ represents the space dimension. 
The basic assumption is that the only relevant
lengthscale is the correlation length $\xi$, and so $\beta f_s \sim \xi^{-D}$.
Consider next a finite-size system where the bulk 
behavior is assumed to
be modified by a function of $L/\xi$, where $L$ is the lattice size, i.e.,
\begin{equation}
\beta f_s = \xi^{-D} {\tilde f}\left({ L \over \xi }\right) \; .
\end{equation}
Thus, we can write the scaling of the total singular
free energy $F_s (= L^D f_s)$ as
\begin{equation}
\beta F_s = {\tilde F}\left({ L \over \xi }\right)
= {\hat F}( L^{1/\nu} (T - T_c) )\; .
\end{equation}
The last expression follows because $\xi \sim (T - T_c)^{-\nu}$ and we
have taken the argument of the function ${\tilde F}$ to the power
$1/\nu$.

As $I_{\rm rms}$ is the derivative of $F_s$ with respect to the twist angle
and since the twist angle is dimensionless, the scaling of $I_{\rm rms}$
is the same as that as of $F_s$, i.e.,
\begin{equation}
\beta I_{\rm rms} = {\hat I}[ L^{1/\nu} (T - T_c) ] \; .
\end{equation}
Note that we have been careful to maintain the factor of $\beta$ in our
analysis. However, if $T_c > 0$, then, since the critical region is close to
$T_c$, we can replace $\beta$ by $1/T_c$ and incorporate this {\em constant}\/
factor into the
scaling function, i.e.,
\begin{equation}
I_{\rm rms} = \tilde{I}[L^{1/\nu}(T - T_c)] \quad \quad (T_c > 0) .
\label{scale_tgt0}
\end{equation}
By contrast, if $T_c=0$ we have to keep the variation in $\beta$ and so
\begin{eqnarray}
I_{\rm rms} & = & T \hat{I}(L^{1/\nu}T)  \nonumber \\
& = & L^{-1/\nu} \tilde{I}(L^{1/\nu}T) \quad \quad (T_c = 0) \; ,
\label{scale_teq0}
\end{eqnarray}
where the scaling functions $\hat{I}$ and $\tilde{I}$ are simply related to
each other.
Equation (\ref{scale_teq0}) indicates that the $T=0$ 
stiffness exponent $\theta$ is
negative and equal to $-1/\nu$. Equation (\ref{scale_tgt0}) 
shows that if $T_c >0$
the curves for $I_{\rm rms}$ for different sizes intersect at the
critical point, whereas Eq.~(\ref{scale_teq0}) shows that if $T_c = 0$, then
the data decrease with increasing size at $T=0$.

According to standard finite-size scaling the spin-glass susceptibility,
defined in Eq.~(\ref{chisg}), behaves as
\begin{equation}
\chisg = L^{2 - \eta}\tilde{\chi}_{_{\mathrm SG}}[L^{1/\nu}(T - T_c)]\; ,
\label{chi_scale}
\end{equation}
which means that at criticality it 
diverges with a power law, i.e.,
\begin{equation}
\chisg \sim L^{2 - \eta} \quad\quad (T = T_c) \; .
\label{chi_tc}
\end{equation}
Equations (\ref{chi_scale}) and (\ref{chi_tc}) are valid whether the
transition is at $T_c=0$ or $T_c > 0$. The power law prefactor in
Eq.~(\ref{chi_scale}) with an unknown exponent complicates the 
analysis of $\chisg$
compared with that for $I_{\rm rms}$ which does not have such a prefactor if 
$T_c >0$, see Eq.~(\ref{scale_tgt0}). Even if $T_c=0$, where a prefactor does
arise [see Eq.~(\ref{scale_teq0})], the exponent in the prefactor is the {\em
same}\/ as that in the argument of the scaling function, so we do not have an
{\em extra}\/ exponent to determine. 

\begin{figure}[tb]
\centerline{\epsfxsize=\columnwidth \epsfbox{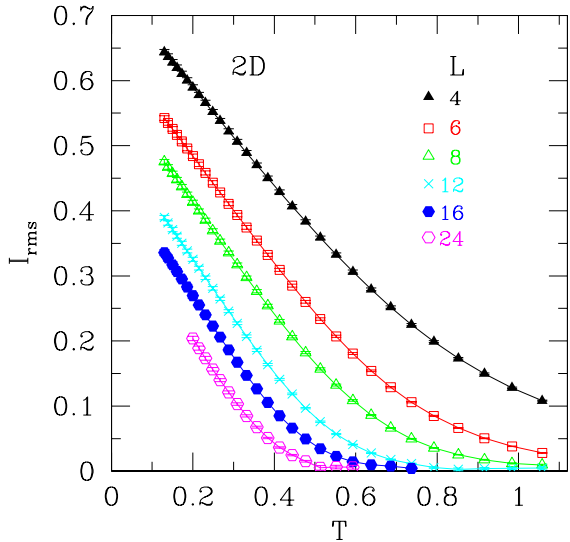}}
\vspace{-1.0cm}
\caption{
Root-mean-square current $I_{\rm rms}$ in two dimensions as a function of
temperature for different system sizes.
At all temperatures the data decrease with increasing $L$ indicating, from
Eq.~(\ref{scale_tgt0}), that if $T_c$ is finite it must be less than $0.13$.
}
\label{irms}
\end{figure}

\begin{figure}
\centerline{\epsfxsize=\columnwidth \epsfbox{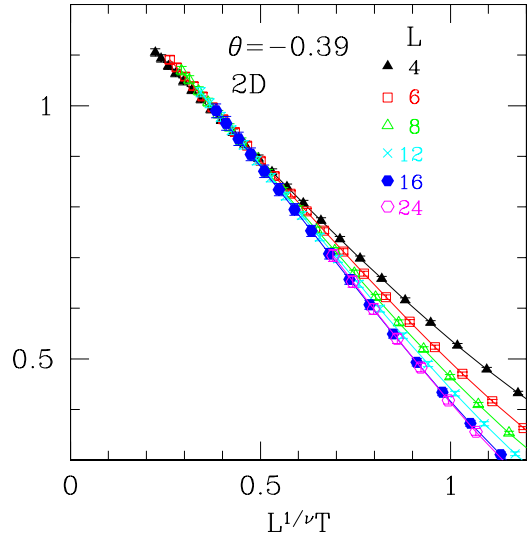}}
\vspace{-1.0cm}
\caption{
Scaling plot of the root-mean-square current $I_{\rm rms}$ in 
two dimensions according to the form expected if $T_c=0$, 
Eq.~(\ref{scale_teq0}). We see acceptable scaling of the data at low
temperatures.
Deviations at higher $T$ are presumably due to corrections to scaling. This
plot is for $\theta \equiv -1/\nu = -0.39$. 
}
\label{irms_scale}
\end{figure}

\section{Results for $D = 2$}
\label{results:2d}

In Fig.~\ref{irms} we present results for $I_{\rm rms}$ in two dimensions.
At all temperatures the data decrease with increasing $L$ indicating, from
Eq.~(\ref{scale_tgt0}), that $T_c$ must be less than the range of temperatures
studied. In fact, we shall find it impossible to scale the data for $I_{\rm
rms}$ for {\em any}\/ finite $T_c$,
in contrast to the results of
Kim\cite{kim:00} and Choi and Park,\cite{choi:99}
who find a finite-temperature transition at $T_c = 0.22$.

In Fig.~\ref{irms_scale} we show a finite-size
scaling plot of the data in Fig.~\ref{irms} according to Eq.~(\ref{scale_teq0}).
Although the data do not scale well over the whole range, the data at low-$T$
{\em do}\/ scale quite well, since the plot shows that
the data collapse if $L^{1/\nu}T$ is
small for all values of $L$, and over the whole range of $L^{1/\nu}T$
for the largest sizes. 
The data in Figs.~\ref{irms} and \ref{irms_scale} are therefore consistent
with a zero temperature transition, but with significant corrections to
scaling at intermediate temperatures. We estimate the stiffness exponent to
be
\begin{equation}
\theta = -0.39 \pm 0.03\;.
\end{equation}
The above error bar is estimated by varying $\theta$ slightly until
the data do not collapse well.
This result is consistent with recent work of Akino and
Kosterlitz\cite{akino:02} who find $\theta = -0.36 \pm 0.01$.

\begin{figure}[tb]
\centerline{\epsfxsize=\columnwidth \epsfbox{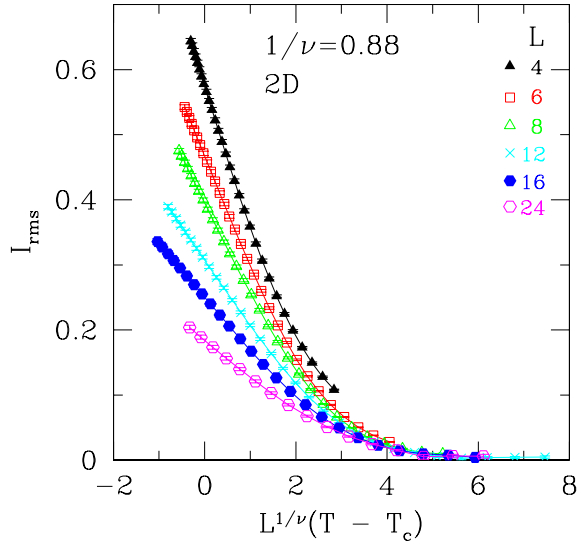}}
\vspace{-1.0cm}
\caption{
Scaling of the current according to Eq.~(\ref{scale_tgt0}) with the parameters
of Choi and Park (Ref.~\onlinecite{choi:99}): $T_c = 0.22$,
and $1/\nu = 0.88$.
One can clearly
see that the scaling is very poor, {\em especially}\/ in the vicinity of the
proposed $T_c$.
}
\label{irms_tc0.22}
\end{figure}

Figure \ref{irms_tc0.22} shows the current
data according to Eq.~(\ref{scale_tgt0})
assuming the parameters found by Choi
and Park\cite{choi:99}: $T_c = 0.22$, $1/\nu = 0.88$.
The 
scaling is {\em very}\/ poor especially near the proposed $T_c$. By contrast,
if the
deviations are due to corrections to scaling, they should be smaller near
$T_c$. Hence our data are incompatible with the claim of
Kim\cite{kim:00} and Choi and Park\cite{choi:99} that
$T_c = 0.22$. In fact, we are unable to get a reasonable fit to the data
for $I_{\rm rms}$
according to Eq.~(\ref{scale_tgt0}) for {\em any}\/ finite $T_c$. 

\begin{figure}[tb]
\centerline{\epsfxsize=\columnwidth \epsfbox{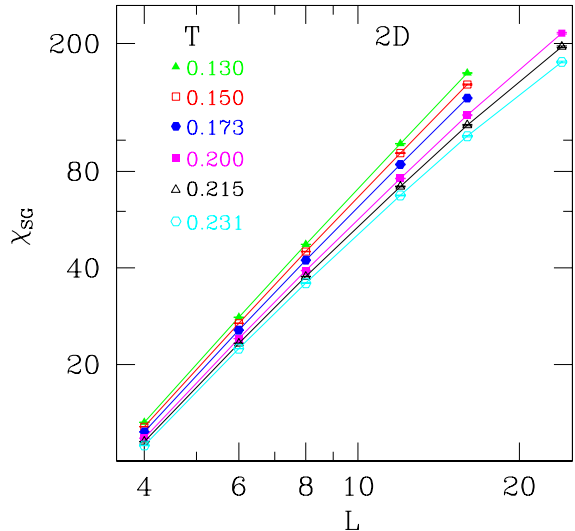}}
\vspace{-1.0cm}
\caption{
Log-log plot of the spin-glass
susceptibility $\chisg$ as a function of system size $L$ for
several different low temperatures. We clearly see a
downward curvature in the data
indicating that $T_c < 0.13$. 
}
\label{logchi}
\end{figure}

Next we present our data for the spin-glass susceptibility $\chisg$.
Figure~\ref{logchi} is a log-log plot of $\chisg$ vs $L$ at
several low temperatures. According to Eq.~(\ref{chi_tc}) 
the data should lie on a straight line at $T_c$. However,
the data in the vicinity of $T = 0.22$, the transition temperature claimed by 
Kim\cite{kim:00} and Choi and Park,\cite{choi:99}
show a strong downward curvature, indicating that this
is actually {\em above}\/ $T_c$. Only around
the lowest temperature where we have
data, $T = 0.13$, is the curvature small, although it still greatly exceeds
the error bars.
This indicates that $T_c < 0.13$, which is
compatible with our data for $I_{\rm rms}$. 

\begin{figure}[tb]
\centerline{\epsfxsize=\columnwidth \epsfbox{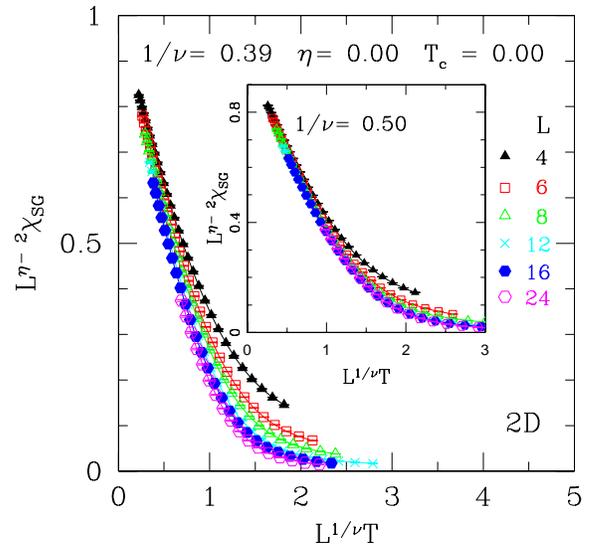}}
\vspace{-1.0cm}
\caption{
Scaling of the spin-glass susceptibility $\chisg$ according to
Eq.~(\ref{chi_scale}) with $T_c = 0$.
The data for large sizes and low temperatures collapse
with $\eta = 0$ and $1/\nu = 0.39$.
The inset shows a scaling plot for the optimal value $1/\nu = 0.50$ (and $\eta
= 0$). For these values of exponents
the collapse extends to a larger range of sizes.
}
\label{chisg_tc0}
\end{figure}

Figure \ref{chisg_tc0} shows a scaling plot according to Eq.~(\ref{chi_scale})
for $T_c = 0$,
and
$1/\nu = 0.39$, the same parameters found in the
scaling of $I_{\rm rms}$, together with $\eta = 0$ which is expected at a
zero-temperature transition in two dimensions.
The data at low temperatures and for the
largest sizes scale well, but the data away from this range show deviations.
Allowing $1/\nu$ to vary we find 
$1/\nu = 0.50 \pm 0.03$. 
The
inset shows data for the optimal value,
where only the $L=4$ and 6 data are not part of the scaling function for
all $L^{1/\nu} T$.
The fact that the best
values of $1/\nu$ are not precisely the same when obtained from $\chisg$ and
$I_{\rm rms}$ presumably indicates that scaling is only valid for fairly low
temperatures and large sizes, and that, despite our working at quite low
temperatures, we have only a limited range of data which are fully in the
scaling regime. 

\begin{figure}[tb]
\centerline{\epsfxsize=\columnwidth \epsfbox{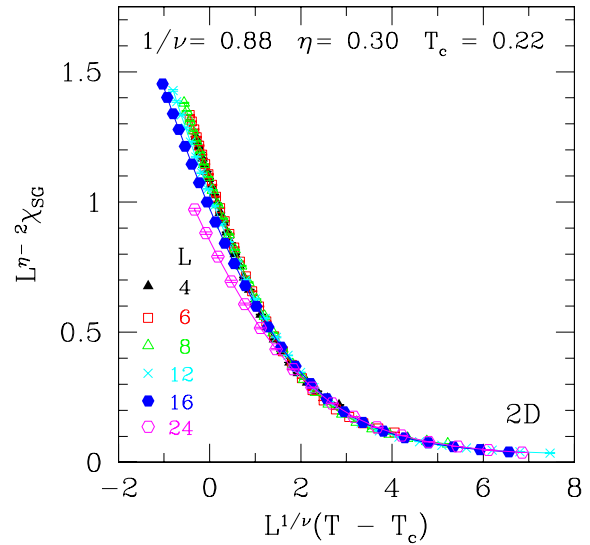}}
\vspace{-1.0cm}
\caption{
Scaling of the spin-glass susceptibility $\chisg$ according to
Eq.~(\ref{chi_scale}) for $T_c = 0.22$, $1/\nu = 0.88$ and $\eta = 0.30$,
as suggested by Choi and Park.\cite{choi:99}
}
\label{chisg_tc0.22}
\end{figure}

\begin{figure}[tb]
\centerline{\epsfxsize=\columnwidth \epsfbox{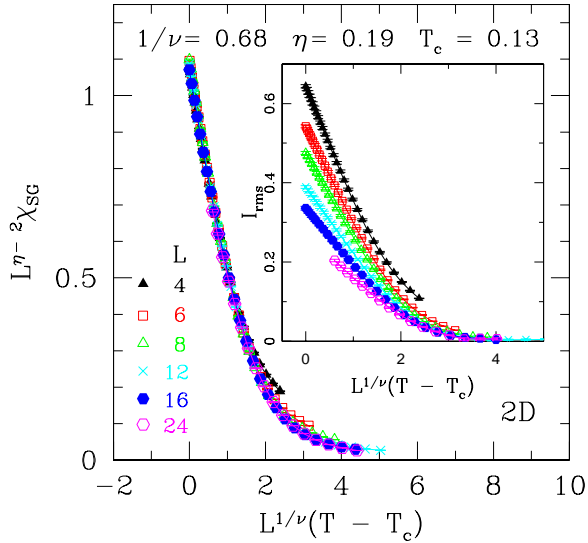}}
\vspace{-1.0cm}
\caption{
Scaling of the spin-glass susceptibility $\chisg$ according to
Eq.~(\ref{chi_scale}) for $T_c = 0.13$, the lowest temperature in our
simulations, $1/\nu = 0.68$ and $\eta = 0.19$. 
The inset shows the scaling of the $I_{\rm rms}$ according to
Eq.~(\ref{scale_tgt0}).
One can clearly see that this data do not scale.
}
\label{chisg_tc0.13}
\end{figure}

In Fig.~\ref{chisg_tc0.22} we show the same data as in Fig.~\ref{chisg_tc0}
scaled with the parameters used by Choi and Park in Ref.~\onlinecite{choi:99}.
We note that the quality of the data collapse is poor
near $T_c$, whereas if the deviations
were due to corrections to scaling, we would expect it to be best in this
region. We therefore do not consider this scaling to be acceptable.
Furthermore, the scaling of $I_{\rm rms}$ shown in
Fig.~\ref{irms_tc0.22} is clearly much worse, indicating that
$I_{\rm rms}$ is much better able to distinguish between a finite $T_c$ and $T_c
= 0$ than is $\chisg$ because of its simpler finite-size scaling form.

This is further illustrated in
Fig.~\ref{chisg_tc0.13} which shows a scaling plot of $\chisg$ in which we
took
$T_c$ to be the lowest temperature in our simulations, i.e.,
$T_c = 0.13$. The best fit with this $T_c$ has $1/\nu = 0.68$ and $\eta =
0.19$, where $\eta$ is determined
by the requirement that the data scale as well as possible at $T_c$.
By eye, the data scale fairly well, though the deviations are actually
significantly greater than the error bars, as we noted in the above
discussion of the unscaled $\chisg$ data at $T=0.13$.
However, the data for $I_{rms}$, scale very badly with $T_c = 0.13$ as can
be seen in the inset of the figure. In fact, as noted above, the data
for $I_{rms}$ do not scale according to Eq.~(\ref{scale_tgt0})
for {\em any}\/ finite $T_c$.

The data for $\chisg$ enable us to determine the exponent $\eta$ as well as
$1/\nu$.
Assuming $T_c = 0$, as obtained from the $I_{\rm rms}$ data, 
we
find $\eta = 0.0 \pm 0.1$ from a
scaling plot for $\chisg$. This is
consistent with $\eta = 0$, as
expected at a $T=0$ transition in two dimensions, at least if the ground state
is not highly degenerate.

\section{Results for $D = 3$}
\label{results:3d}

The critical region of the three-dimensional 
gauge glass has been investigated in detail by Olson and
Young.\cite{olson:00} In their work they obtain a lower bound for the
stiffness exponent of $\theta \ge 0.18$. 
Some previous results\cite{reger:91,gingras:92,kosterlitz:97,maucourt:97} 
find $\theta$ in the range $0 \le \theta \le 0.077$ whereas
others\cite{akino:02,cieplak:92,moore:94,kosterlitz:98}
find a much larger value,
$0.26 \le \theta \le 0.31$.

\begin{figure}[tb]
\centerline{\epsfxsize=\columnwidth \epsfbox{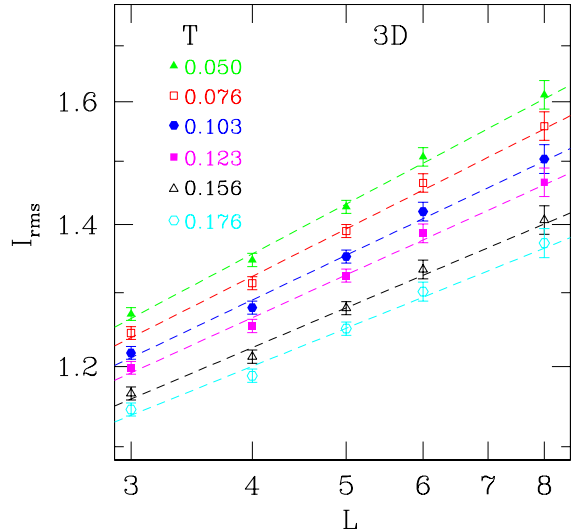}}
\vspace{-1.0cm}
\caption{
Log-log plot of $I_{\rm rms}$ vs $L$ for different low temperatures.
The data fit well to $aL^{\theta}$ with fitting probabilities between
$0.57$ for the lowest temperature $T = 0.050$ and $0.35$ for $T = 0.176$.
}
\label{logi}
\end{figure}

\begin{figure}[tb]
\centerline{\epsfxsize=\columnwidth \epsfbox{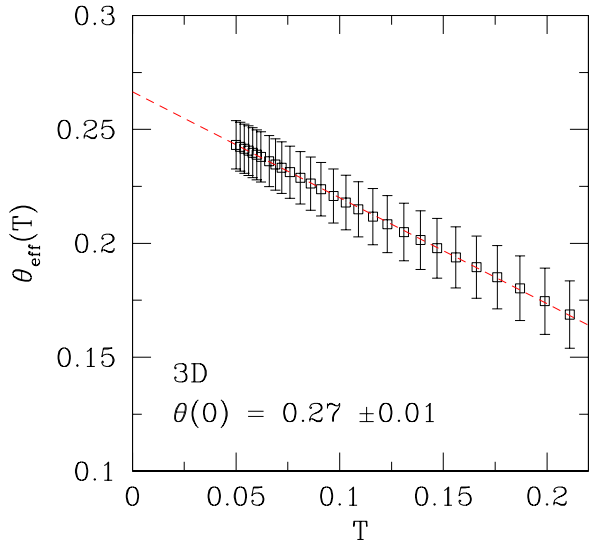}}
\vspace{-1.0cm}
\caption{
Effective stiffness exponent $\theta_{\rm eff}(T)$ as a function of
temperature in three dimensions for low temperatures. The data extrapolate
linearly to $T = 0$.
}
\label{theta}
\end{figure}

We can estimate $\theta$ from our data for $I_{\rm rms}$ since
$I_{\rm rms} \sim L^\theta$ when $L^{1/\nu}(T - T_c)$,
the argument of the scaling function in
Eq.~(\ref{scale_tgt0}), tends to infinity. The data are shown in
Fig.~\ref{logi}, in which one observes that the
slopes in the log-log plot vary somewhat 
with temperature.
Asymptotically, however, one expects the same $L^\theta$ behavior for {\em all}\/ $T <
T_c$, so the deviations from a constant slope presumably indicate that that
the data are not yet at sufficiently large sizes and low enough temperature to
be in the asymptotic region.  
To obtain an estimate of $\theta$ we therefore perform a linear least-squares fit
of
$\ln I_{\rm rms}$ against $\ln L$
for each temperature in order to
obtain an {\em effective}\/ stiffness exponent $\theta_{\rm eff}(T)$ which
depends on the temperature.
Figure~\ref{theta} shows that $\theta_{\rm eff}(T)$
can be fitted well to a linear form
at low temperatures. Extrapolating to $T = 0$ we obtain
\begin{equation}
\theta = 0.27 \pm 0.01
\end{equation}
which is clearly positive and consistent with the results
of Refs.~\onlinecite{olson:00,akino:02,cieplak:92,moore:94,kosterlitz:98}.

\section{Conclusions}
\label{conclusions}

We have shown results from Monte Carlo simulations at low temperatures
of the two- and three-dimensional gauge glass. In two dimensions our
data for $I_{\rm rms}$
are consistent with a zero-temperature transition with $\theta = -0.39
\pm 0.03$. However, it is necessary to go to quite low
temperatures and large sizes to see the expected scaling.
Our data are incompatible with the prediction
made by Kim\cite{kim:00} and Choi and Park\cite{choi:99}
that $T_c = 0.22$. The error bar we quote is purely statistical and systematic
corrections to scaling can increase this. Indeed, our best estimate for
$1/\nu \equiv -\theta$ from the $\chisg$ data is $0.50 \pm 0.03$, which
differs from the value of $1/\nu$ from  $I_{\rm rms}$ by more than the
(statistical) error bars.
However, our claim that $T_c = 0$ is very robust and is not affected by
corrections to finite-size scaling. 

In three dimensions we report
the first reliable estimate of the stiffness exponent
from finite-temperature
Monte Carlo simulations. Monte Carlo has the advantage over ground state methods
that it computes the stiffness directly and is free
from the difficulty of determining the optimal boundary condition changes,
which has been quite controversial for the ground state
approach.
We find
$\theta = 0.27 \pm 0.01$, which agrees with the results of
Refs.~\onlinecite{olson:00,akino:02,cieplak:92,moore:94,kosterlitz:98}.

\begin{acknowledgments}
We would like to thank C.~Dasgupta, R.~T.~Scalettar, P.~Sengupta and A.~Slepoy
for fruitful discussions, and  R.~T.~Scalettar and T.~Olson for comments on an
earlier
version of the manuscript.
HGK acknowledges support from the National Science Foundation under
Grant No.~DMR 9985978. APY acknowledges support from the National Science
Foundation under Grant No.~DMR 0086287, and the EPSRC
under Grant No.~GR/R37869/01. He would also like to thank David Sherrington for
hospitality during his stay at Oxford. 
This research was supported in part by NSF cooperative agreement ACI-9619020
through computing resources provided by the National Partnership for Advanced 
Computational Infrastructure at the San Diego Supercomputer Center and the
Texas Advanced Computing Center. We would like to thank the University of New
Mexico for access to their Albuquerque High Performance Computing Center. 
This work utilized the UNM-Alliance Los Lobos Supercluster. We would also
like to acknowledge computer time on the UC Davis Undergraduate Computing
Cluster and the UCSC Physics Graduate Computing Cluster funded by the
Department of Education Graduate Assistance in the Areas of National Need
program.

\end{acknowledgments}

\bibliography{refs}

\end{document}